\documentclass[aps,prl,preprintnumbers,amsmath,amssymb,latexsym,nofootinbib,array,enumerate,letter,twocolumn,superscriptaddress]{revtex4-1}

\usepackage{subfigure}
\usepackage{cancel}
\usepackage{here}
\usepackage{comment,braket}
\usepackage{epsf}
\usepackage{amsmath}
\usepackage{graphics}
\usepackage{amsfonts}
\usepackage{amssymb}
\usepackage{latexsym}
\usepackage{color}
\usepackage{ulem}
\input{colordvi.tex}
\usepackage{feynmp}
\usepackage{caption}

\usepackage[pdftex]{graphicx}
\usepackage{bm}
\usepackage[pdftex]{hyperref}
\usepackage[svgnames]{xcolor}
\definecolor{phthaloblue}{rgb}{0.0, 0.06, 0.54}
\hypersetup{
    colorlinks=true,
    linkcolor=phthaloblue,
    citecolor=blue,
    filecolor=blue,
    urlcolor=phthaloblue,
    }

\newcommand{\beq}{\begin{eqnarray}} 
\newcommand{\eeq}{\end{eqnarray}}

\def\({\left(}
\def\){\right)}
\def\[{\left[}
\def\]{\right]}

\newcommand{\bel}[1] {\begin{equation}\label{#1}}
\newcommand{\beal}[1] {\begin{eqnarray}\label{#1}}
\newcommand{\be}{\begin{equation}}
\newcommand{\ee}{\end{equation}}
\newcommand{\bea}{\begin{array}} 
\newcommand{\eea}{\end{array}}

\begin{document}
    
\title{Affleck-Dine Dirac Leptogenesis}

\author{Neil D. Barrie}
\email{nlbarrie@gmail.com}
\affiliation{Sydney Consortium for Particle Physics and Cosmology, \\ School of Physics, The University of Sydney, NSW 2006, Australia.}
	
\author{Chengcheng Han}
\email{hanchch@mail.sysu.edu.cn}
\affiliation{School of Physics, Sun Yat-Sen University, Guangzhou 510275, China}
\affiliation {Asia Pacific Center for Theoretical Physics, Pohang 37673, Korea}

\begin{abstract}
We present a minimal framework that realises successful Dirac Leptogenesis through the Affleck-Dine mechanism. A single right-handed neutrino and a neutrinophillic Higgs doublet are introduced to the Standard Model, which couple via a Yukawa interaction. The inflationary setting is induced by a combination of the two Higgs doublets, with their global symmetry violating interactions leading to a net charge generation via the Affleck-Dine mechanism. This simple Standard Model extension exhibits a unique and connected set of phenomenological implications including the resultant baryon asymmetry, inflationary predictions, cosmological implications, relic right-handed neutrinos, and its low energy phenomenology, while also being able to be embedded in various neutrino mass generating mechanisms.
\end{abstract}

\maketitle

\noindent \textbf{Introduction - } 
The existence of the matter-antimatter asymmetry in our universe is an essential piece of evidence for physics beyond the Standard Model (SM). One intriguing solution is Leptogenesis, which illuminates this mystery through exploring the current unknowns of the neutrino sector \cite{Minkowski:1977sc, Yanagida:1979as, Glashow:1979nm, GellMann:1980vs,Fukugita:1986hr}. In the standard thermal Leptogenesis scenario~\cite{Fukugita:1986hr}, an asymmetry is first generated in the lepton sector through the $ \mathcal{CP} $ violating decay of introduced heavy right-handed (RH) neutrinos, before this lepton number is redistributed into baryons by equilibrium sphaleron processes \cite{Kuzmin:1985mm}. 
However, the mass scale of these RH neutrinos is typically required to be higher than  $ 10^7 $ GeV to be consistent with current observations of the baryon asymmetry~\cite{Giudice:2003jh}.\footnote{If the masses of two lightest RH neutrinos are nearly degenerate, the required mass scale can be reduced \cite{Pilaftsis:2003gt, Pilaftsis:2005rv, Drewes:2013gca}.} The high new physics scales associated with these models render terrestrial experimental tests infeasible for the foreseeable future.
F
The introduction of a RH neutrino to the SM opens up alternative possibilities for elucidating the origin of the observed baryon asymmetry. One intriguing prospect is Dirac-genesis~\cite{Dick:1999je, Murayama:2002je}, a scenario in which at least  one of the neutrinos exhibits a Dirac nature. In such models, successful Leptogenesis occurs despite the total $B-L$ of the universe being zero. This is achieved by an equal and opposite lepton number being sequestered from the SM sector into a density of decoupled RH neutrinos prior to the Electroweak Phase Transition (EWPT). Subsequently, the residual lepton asymmetry in the SM sector is transferred into a baryon asymmetry via equilibrium sphaleron processes as in usual Leptogenesis. 

In this letter, we present a straightforward and simple framework in which to realise Dirac-genesis through the introduction of only one additional Higgs doublet and a single RH neutrino. The additional Higgs doublet establishes connections with both the left-handed leptons and the right-handed neutrino through a Yukawa coupling - while preserving $\mathcal{CP}$ and $L$ conservation in the model \cite{Balaji:2004iq,Gabriel:2006ns,Davidson:2009ha,Haba:2011ra,Heeck:2013vha,Machado:2015sha,Gu:2019yvw,Li:2022yna}. An asymmetry in the Higgs sector is generated through the Affleck-Dine mechanism \cite{Affleck:1984fy,Barrie:2021mwi,Barrie:2022cub}, wherein the required initial displaced vacuum value is achieved by configuring the Higgs fields to be the inflaton that drives inflation \cite{Brout:1977ix, Starobinsky:1980te, Sato:1980yn, Guth:1980zm, Linde:1981mu, Albrecht:1982mp,Whitt:1984pd,Jakubiec:1988ef,Maeda:1988ab,Barrow:1988xh,Faulkner:2006ub,Bezrukov:2011gp}.  

The possibility of Affleck-Dine Dirac Leptogenesis has been suggested in previous literature \cite{Balaji:2004iq,Abel:2006hr,Senami:2007up}. However, a complete renormalizable model that satisfies the necessary conditions for a viable Affleck-Dine mechanism, without supersymmetry, has not been presented. Typically, the Affleck-Dine mechanism is applied within the framework of supersymmetry, benefiting from the existence of a flat direction in the potential through which a large field vacuum can be achieved during inflation. The challenge in the non-supersymmetric Affleck-Dine mechanism lies in the absence of such flat directions for the new scalar field. One solution is to consider the Affleck-Dine field as the Nambu pseudo-goldstone boson of a spontaneously global $U(1)$ symmetry, ensuring the flatness of the potential through its inherent shift symmetry~\cite{Harigaya:2019emn, Co:2019wyp, Berbig:2023uzs}. Another approach involves considering the Affleck-Dine field as the inflaton triggering inflation~ \cite{Hertzberg:2013mba, Lozanov:2014zfa, Yamada:2015xyr, Bamba:2016vjs,Bamba:2018bwl, Cline:2019fxx,Barrie:2020hiu, Lin:2020lmr, Kawasaki:2020xyf, Kusenko:2014lra, Wu:2019ohx, Charng:2008ke, Ferreira:2017ynu,Bettoni:2018utf, Babichev:2018sia, Rodrigues:2020dod, Lee:2020yaj, Enomoto:2020lpf, Lloyd-Stubbs:2020sed, Mohapatra:2021aig, Mohapatra:2022ngo,Barrie:2021mwi,Barrie:2022ake,Barrie:2022cub,Han:2022ssz,Borah:2022qln,Han:2023vme,Han:2023kjg}, requiring a non-minimal coupling of the Affleck-Dine field to gravity. Note that this model only contains an additional Higgs doublet and a RH neutrino. As far as we know, this is the first successful realisation of the Affleck-Dine mechanism within a minimal particle extension of the SM, without the introduction of higher dimension operators.\\


\noindent \textbf{The Affleck-Dine Mechanism - }
The observed baryon asymmetry of the universe has been measured to be,
\begin{equation}
	\eta_B = \frac{n_B}{s} \simeq 8.5\times 10^{-11}~,
	\label{eta1}
\end{equation}
where $n_B$ and $s$ are the baryon number and entropy densities of the universe, respectively \cite{Aghanim:2018eyx}. The Affleck-Dine mechanism was proposed to explain this asymmetry through the generation of angular motion in the phase of a complex scalar field $\phi$, which is charged under a global $U(1)$ symmetry \cite{Affleck:1984fy}. The $\phi$ must acquire a large initial field value in the early universe, which subsequently begins to oscillate once the Hubble parameter falls below its mass $m$.  Assuming that the scalar potential $ V $ contains an explicit $U(1)$ breaking term, the generated angular motion will correspond to a net $U(1)$ charge asymmetry with number density,
\begin{eqnarray}
n_Q = 2 Q\textrm{Im} [\phi^\dagger \dot \phi]=  Q \chi^2 \dot{\theta }~,
\end{eqnarray}
where  $\phi= \frac{1}{\sqrt{2}}\chi e^{i\theta}$. Thus, to obtain a non-zero $n_Q$, we require motion in the complex phase $\theta$ and for $\chi$ to have a non-zero vacuum value. Directed motion will be generated through the $U(1)$ breaking terms in the scalar sector, while the $U(1)$ conserving terms will be dominant during the inflationary epoch and source the non-zero vacuum value.

If the $U(1)$ symmetry is related to the global $U(1)_B$ or $U(1)_L$ symmetries, this mechanism can generate an asymmetry prior to the EWPT that will be redistributed by equilibrium sphalerons. In the following, we will take the $\phi$ field to be a mixed state of the SM Higgs and a new neutrinophillic Higgs doublet, which exhibits a complex phase associated with a global $U(1)_X$ symmetry.\\


\noindent \textbf{Two Higgs Doublet Model - } 
To realise the Affleck-Dine mechanism during inflation and successful Leptogenesis, we consider a model consisting of an additional neutrinophillic Higgs doublet $\Phi_2= (\Phi^0_2,~ \Phi^-_2)$, and a RH neutrino $\nu_R$ - besides the SM particle content. The Lagrangian, including the SM-like Higgs $\Phi_1=(\Phi^0_1, ~\Phi^-_1)$, is given by \cite{Gabriel:2006ns,Davidson:2009ha,Machado:2015sha,Balaji:2004iq,Haba:2011ra,Heeck:2013vha,Gu:2019yvw}, 
\begin{eqnarray}
\frac{\mathcal L}{\sqrt{-g}}&=&-\frac{1}{2} M_p^2 R -f(\Phi_1,\Phi_2) R -g^{\mu\nu} (D_\mu \Phi_1)^\dagger (D_\nu \Phi_1) \nonumber \\
&& -g^{\mu\nu} (D_\mu \Phi_2)^\dagger (D_\nu \Phi_2) -V(\Phi_1, \Phi_2) + \mathcal L_{\textrm{Yuk}}~, \label{Lagrange1B}
\end{eqnarray}
where
\begin{eqnarray}
\mathcal L_{\textrm{Yuk}} =    \mathcal L^{\rm SM}_{\textrm{Yuk}} + y \bar L  \Phi_2 \nu_R~  + h.c.~,
\end{eqnarray}
and
\begin{align}
V(\Phi_1,\Phi_2)&=m_{11}^2\Phi_1^\dagger\Phi_1 +m_{22}^2\Phi_2^\dagger\Phi_2+\lambda_3 \Phi_1^\dagger\Phi_1 \Phi_2^\dagger\Phi_2\nonumber\\
& +\dfrac{\lambda_1}{2}(\Phi_1^\dagger \Phi_1)^2+\dfrac{\lambda_2}{2}(\Phi_2^\dagger \Phi_2)^2
+\lambda_4 \Phi_1^\dagger\Phi_2 \Phi_2^\dagger\Phi_1\nonumber\\
&+(\dfrac{\lambda_5}{2}(\Phi_1^\dagger\Phi_2)^2 + \textrm{h.c.})~.
\end{align}

The interactions in this Lagrangian respect a global $U(1)_L$ symmetry, we call lepton number, for which the $\nu_R$ is chosen to have charge $+1$, thereby forbidding a Majorana mass term. There exists another global symmetry, $U(1)_X$, under which $\Phi_2$ carries a charge of $+1$ and $\nu_R$ a charge of $-1$. Importantly, this symmetry is broken by the $\lambda_5$ coupling term between the two Higgs doublets. There remains an accidental $Z_2$ symmetry even in the presence of the $\lambda_5$ term. There are also possible $\lambda_6$ and $\lambda_7$ terms in a general two Higgs doublet model, such terms violate $\mathcal{CP}$ and can be safely removed by imposing  $\mathcal{CP}$ symmetry. 

Due to the presence of the residual $Z_2$ symmetry, the coupling between $\nu_R$ and the SM Higgs is prohibited and the coupling of $\Phi_2$ with SM quarks is also forbidden - leading to a neutrinophillic two Higgs Doublet model \cite{Gabriel:2006ns,Davidson:2009ha,Machado:2015sha}. Simultaneously, the possible $ m_{12}^2(\Phi_1^\dagger\Phi_2) $ term is disallowed due to the residual $Z_2$ symmetry. The absence of the $m_{12}$ term is crucial to prevent the mixing of $\Phi_2$ with the SM Higgs, safeguarding against the washing out of the chemical potential of ${\Phi_2}$ in the early universe. In this scenario, the new Higgs doublet $\Phi_2$ does not acquire any vacuum value, akin to the inert Higgs doublet model \cite{Deshpande:1977rw,Ma:2006km,Belyaev:2016lok}. 

The $ U(1)_X $ breaking term is essential to the generation of a $U(1)_X$ asymmetry through the Affleck-Dine mechanism during inflation. After reheating and the subsequent decoupling of $\Phi_2$, the entirety of the generated $X$ asymmetry is sequestered by the RH neutrino through the associated Yukawa coupling. This corresponds to a net lepton asymmetry in the RH neutrino sector which results in a corresponding but oppositely charged lepton asymmetry in the SM sector - maintaining the conservation of the total lepton number. The lepton asymmetry held in the SM sector as a net $ B-L $ is then redistributed by sphalerons in the known way, leading to a net baryon asymmetry.\\


\noindent \textbf{The Inflationary Trajectory - }
The dynamics of the inflationary setting are integral to successful Leptogenesis in this scenario, and will be induced by a mixture of the two Higgs' doublets. To achieve this, the two doublets will have non-minimal couplings to gravity, which lead to the flattening of the scalar potential at large field values. This framework acts analogously to standard Higgs inflation, and results in a Starobinsky-like inflationary scenario \cite{Starobinsky:1980te,Whitt:1984pd,Jakubiec:1988ef,Maeda:1988ab,Barrow:1988xh,Faulkner:2006ub,Bezrukov:2007ep,Bezrukov:2008ut,GarciaBellido:2008ab,Barbon:2009ya,Barvinsky:2009fy,Bezrukov:2009db,Giudice:2010ka,Bezrukov:2010jz,Burgess:2010zq,Lebedev:2011aq,Bezrukov:2011gp,Lee:2018esk,Choi:2019osi}.

The non-minimal couplings are of the following form, 
\begin{eqnarray}
f(H,\Delta) = \xi_1 |\Phi_1|^2 + \xi_2 |\Phi_2|^2=\frac{1}{2}\xi_1\rho^2_1+\frac{1}{2} \xi_2 \rho^2_2~,
\end{eqnarray}
where we have utilized the polar coordinate parametrization  $\Phi_1 \equiv (\frac{1}{\sqrt{2}} \rho_1 e^{i\eta}, 0)$,  $\Phi_2 \equiv (\frac{1}{\sqrt{2}}\rho_2 e^{i\theta},0)$ for the direction of inflation. Interestingly, a unique inflationary trajectory appears when considering a scenario with two non-minimally coupled scalars \cite{Lebedev:2011aq}, with the ratio of the two scalars being fixed in the large field limit,
\begin{eqnarray}
\frac{ \rho_1}{\rho_2}\equiv \tan \alpha = \sqrt{\frac{\lambda_2\xi_1-(\lambda_{3}+\lambda_{4})\xi_2}{ \lambda_1\xi_2 -(\lambda_{3} +\lambda_{4})\xi_1}}~.
\end{eqnarray}
To ensure this is the trajectory, we require $\lambda_2\xi_1-(\lambda_{3}+\lambda_{4})\xi_2>0$ and $2\lambda_1\xi_2 -(\lambda_{3}+\lambda_{4}) \xi_1>0$. 

The inflationary setting is then an approximately single field scenario with the inflaton defined as $ \varphi $,
\begin{eqnarray}
&& \rho_{1}=\varphi\sin\alpha,~\rho_{2} = \varphi\cos\alpha,~  \nonumber \\
&& \xi \equiv \xi_1 \sin^2\alpha +\xi_2\cos^2\alpha~,
\end{eqnarray}
with the Lagrangian becoming,
\begin{eqnarray}
\frac{\mathcal L}{\sqrt{-g}} &=& -\frac{1}{2} M_p^2 R -\frac{1}{2} \xi  \varphi^2  R  - \frac{1}{2} g^{\mu\nu} \partial_\mu \varphi \partial_\nu \varphi  \nonumber \\
&& -\frac{1}{2} \varphi^2 \cos^2\alpha~ g^{\mu\nu} \partial_\mu \theta \partial_\nu \theta-V(\varphi, \theta) ~,
\label{lagrang1B}
\end{eqnarray}
where 
\begin{eqnarray}
\hspace{-0.5cm}V(\varphi, \theta)  = \frac{m^2}{2}\varphi^2 + \frac{1}{4}\lambda \varphi^4  + 2 \tilde \lambda_5 \cos 2\theta \varphi^4 ~,
\end{eqnarray}
with $\lambda=2(\lambda_1\sin^4\alpha+\lambda_2\cos^4\alpha)+2(\lambda_3+\lambda_4)\sin^2\alpha\cos^2\alpha$, $ \tilde \lambda_5=\frac{\lambda_5}{8} \sin^2\alpha\cos^2\alpha$, and $m^2=2 m_{11}^2+2 m_{22}^2$.
The $\theta$ is taken as a dynamical field, given that its motion generates the $ X $ charge asymmetry. In the large field limit, during inflation, the potential is dominated by the $ \lambda $ term. 

We translate the Lagrangian in Eq. (\ref{lagrang1B}) from the Jordan frame to the Einstein frame, through the transformations, $ \tilde{g}_{\mu \nu} = \Omega^2 g_{\mu\nu}$ and $ \Omega^2= 1+ \xi \varphi^2 /M^2_p~$, and by reparametrizing $\varphi$ in terms of the canonically normalized scalar $\chi$ \cite{Wald:1984rg,Faraoni:1998qx}. Resulting in the Einstein frame Lagrangian,
\begin{eqnarray}
\frac{\mathcal L}{\sqrt{-g}} =-\frac{M_p^2}{2}  R - \frac{1}{2}  g^{\mu\nu} \partial_\mu \chi\partial_\nu \chi \nonumber \\-
\frac{1}{2} f(\chi)   g^{\mu\nu} \partial_\mu \theta \partial_\nu\theta- U(\chi,\theta) ~,
\label{L1a}
\end{eqnarray}
where 
\begin{eqnarray}
&& \hspace{-0.5cm} f(\chi) \equiv   \frac{\varphi(\chi)^2 \cos^2 \alpha}{\Omega^2(\chi)},  ~\textrm{and}~ U(\chi,\theta)  \equiv   \frac{V(\varphi(\chi), \theta)}{\Omega^4(\chi)}~,
\label{potential1a}
\end{eqnarray}
with the large field limit of the $ \chi $ potential reproducing that of Starobinsky inflation,
\begin{eqnarray}
U_{\textrm{inf}}(\chi)= \frac{3}{4}m_S^2 M_p^2 \left(1-e^{-\sqrt{\frac{2}{3}}\frac{\chi}{Mp}}\right)^2~,
\end{eqnarray}
where $ m_S=\sqrt{\frac{\lambda M_p^2}{3\xi^2}}\simeq 3 \cdot 10^{13} $ GeV \cite{Faulkner:2006ub,Akrami:2018odb}. Thus, the equations of motion are,
\begin{eqnarray}
&& \ddot{\chi} -\frac{1}{2} f'(\chi)  \dot \theta^2 + 3 H \dot \chi + U_{,\chi} =0  ~,\nonumber \\
&& \ddot{\theta} + \frac{f'(\chi)}{f(\chi)}  \dot \theta \dot \chi + 3 H \dot \theta +\frac{1}{f(\chi)} U_{,\theta} =0 ~.
\label{thetaEOM}
\end{eqnarray}

\begin{figure}

\includegraphics[width=0.75\columnwidth]{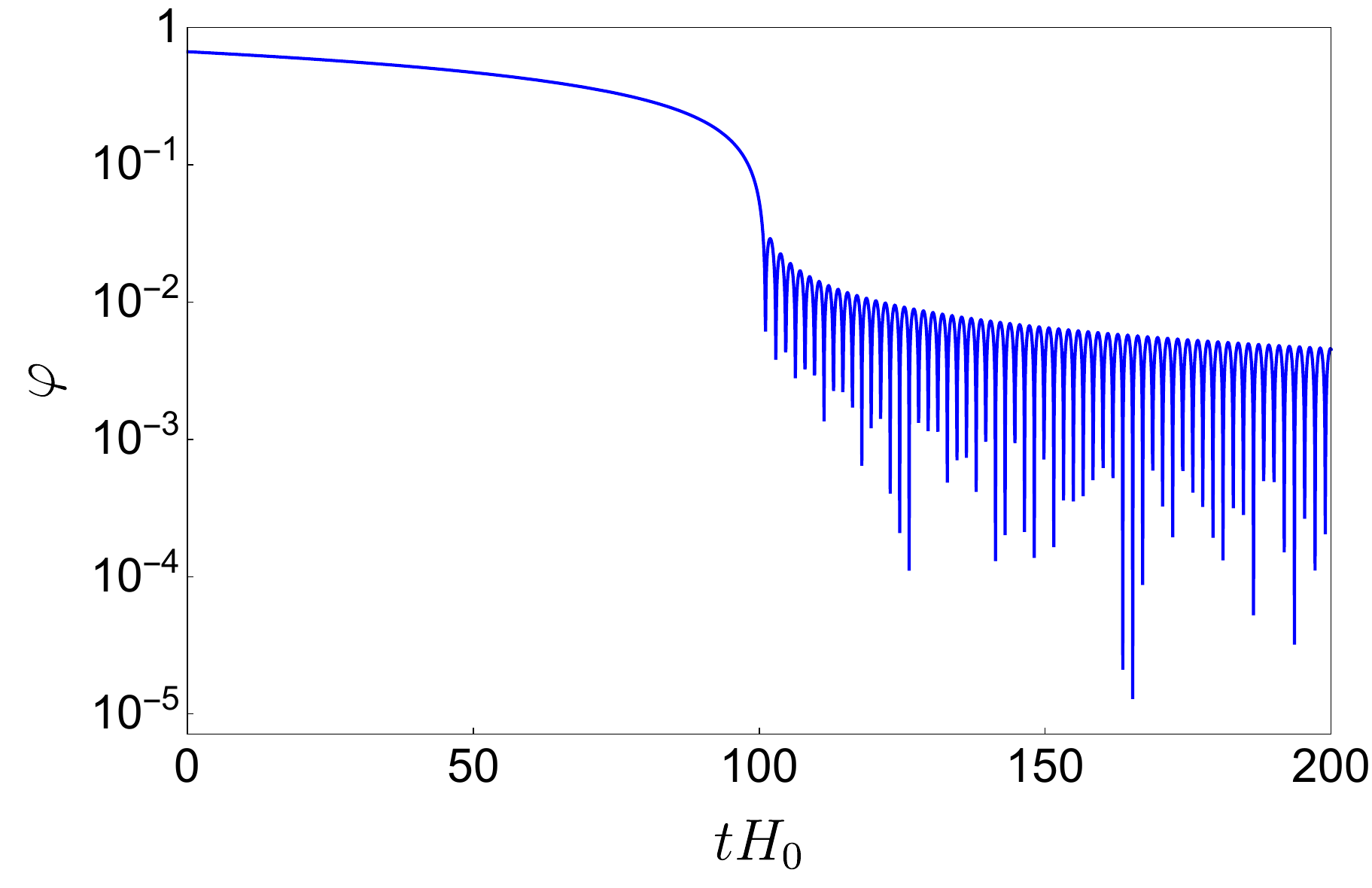}
\caption{The evolution of $ \varphi $ throughout the inflationary and the oscillatory periods before reheating, where we use Planck units and define $H_0= m_S/2$, while fixing $\xi=300$ and $\lambda=4.5\cdot 10^{-5}$.}
	\label{phi} 
\end{figure}

The inflationary trajectory is negligibly affected by the dynamics of $ \theta $ due to the smallness of the $\lambda_5$ term. Thus, the inflationary observables in this scenario are consistent with the Starobinsky model, and in excellent agreement with current observational constraints  \cite{Akrami:2018odb}. The evolution of the field $ \varphi $ during inflation and the oscillatory period before reheating are depicted in Figure \ref{phi}.

\begin{figure}
	\includegraphics[width=0.8\columnwidth]{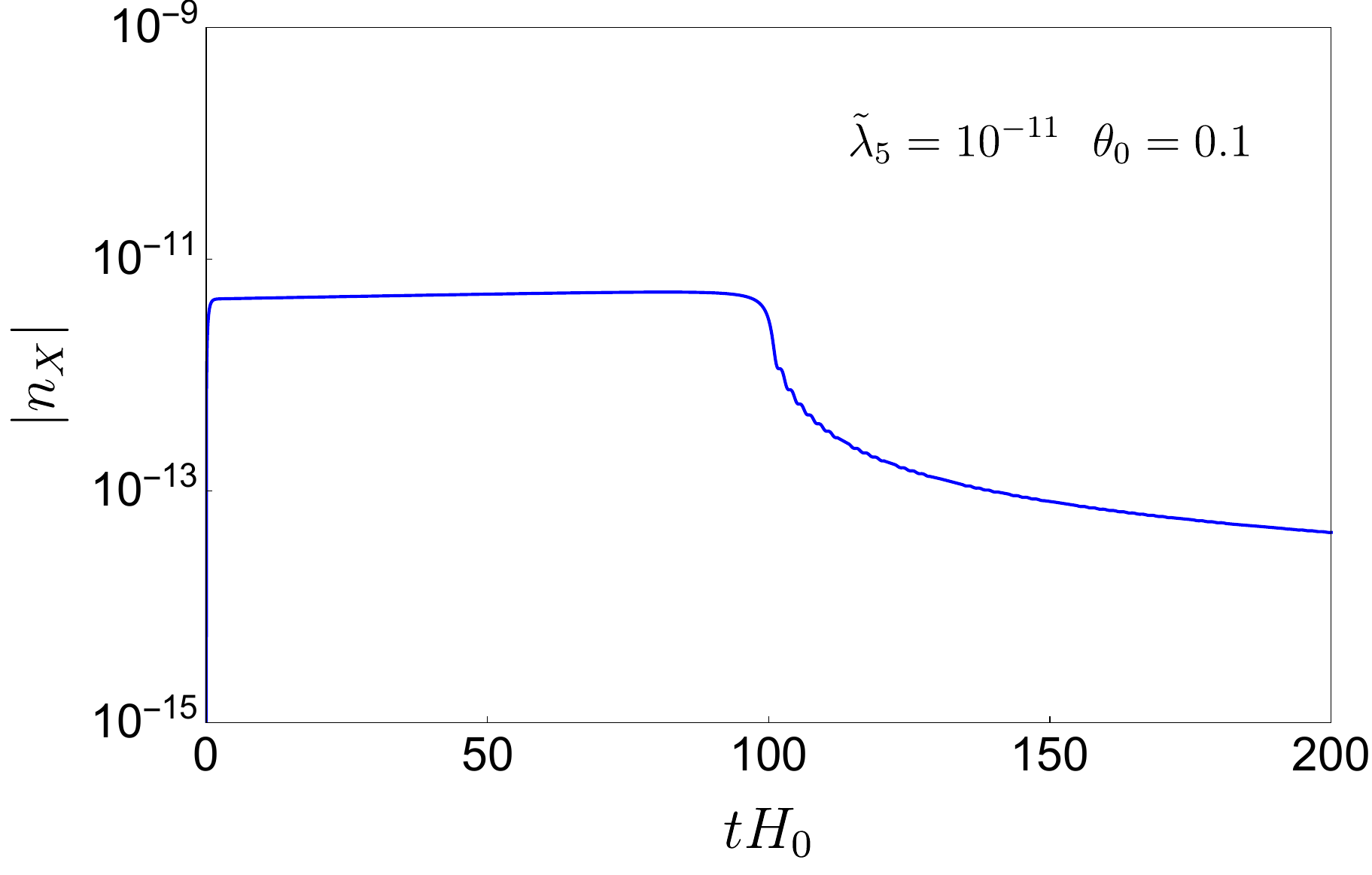}
	\caption{The generated $ n_X $ throughout inflationary and the oscillatory periods prior to reheating, where we use Planck units and define $H_0= m_S/2$, while fixing $\xi=300$ and $\lambda=4.5\cdot 10^{-5}$.}
	\label{nX} 
\end{figure}

The Affleck-Dine mechanism is realised during this inflationary setting, and the generated $ X $ asymmetry at the end of inflation may be estimated from the known slow roll evolution of both $\chi$ and $\theta$,
\begin{eqnarray}
\dot{\chi} \simeq -\frac{M_p U_{,\chi}}{\sqrt{3U}}, ~~\textrm{and}~~ \dot{\theta} \simeq -\frac{M_p U_{,\theta}}{f(\chi)\sqrt{3U}} ~.
\label{dottheta_inf11}
\end{eqnarray}
from which we may determine the $ n_X $ generated during inflation,
\begin{eqnarray}
n_X^{\textrm{inf}} &&=  Q_X f(\chi) \dot \theta  \simeq \frac{8\tilde{\lambda}_5 \sin(2\theta)}{\sqrt{3\lambda}\xi} M_p^3.
\label{asym}
\end{eqnarray}
which is of order of $10^{-12} M_p^3$ for $\tilde{\lambda}_5 = 10^{-11}$ and $ \theta_0=0.1 $, as depicted in Figure \ref{nX}. 

The exact value of $\eta_B$ generated by the end of the EWPT is sensitive to the reheating process. A good estimate can be obtained by using the expected reheating temperature in usual Higgs inflation ($ T_{\textrm{reh}} \sim 2.2\cdot10^{14} $ GeV) ~\cite{Garcia-Bellido:2008ycs, Bezrukov:2008ut, Ema:2016dny, DeCross:2015uza, DeCross:2016cbs, DeCross:2016fdz, He:2018mgb,He:2020ivk,He:2020qcb, Sfakianakis:2018lzf}, which leads to the requirement of  an $n_X$ of order $10^{-18} M_p^{3}$  at the end of inflation to explain the observed $\eta_B$.

Once inflation ends, the generated charge density follows an approximate matter-like dilution during the oscillatory epoch prior to reheating, and the onset of the radiation dominated epoch.\\


\noindent \textbf{Leptogenesis via $ \nu_R $ and Washout Processes - }
We have just shown how to generate a net $ X $ charge density during inflation, so now we must determine how this translates into the observed matter-antimatter asymmetry. To calculate the final baryon asymmetry derived from the $ X $ charge asymmetry, we must carefully consider possible washout processes prior to the EWPT. 

The key feature of the Dirac Leptogenesis scenario is the success of Leptogenesis within a lepton number conserving model. This will occur through the sequestering of a net lepton asymmetry into the decoupled RH neutrino sector, which has a corresponding equal and opposite net lepton asymmetry in the SM sector which is redistributed by sphalerons. In our case, this sequestering occurs once the $\Phi_2$ decouples from the thermal plasma. Due to $U(1)_X$ conservation after reheating and the decoupling of $\Phi_2$,  the entirety of the generated $X$ asymmetry will be contained within the RH neutrinos through the associated Yukawa coupling. Thus, the lepton asymmetry in the SM sector prior to sphaleron redistribution is exactly given by $n_L= -n_X$~.

All the interactions in this scenario also conserve the $B-L$ symmetry, and thus, via electroweak sphaleron redistribution, the resultant baryon asymmetry is given by,
\begin{eqnarray}
\eta_B= \frac{28}{79}\frac{n_X}{s}~.
\end{eqnarray}
this relation in combination with Eqs. (\ref{asym}) and (\ref{eta1}), allow the parameters that lead to successful Leptogenesis to be deduced. Ample parameter space is available, while generically requiring a small $\lambda_5$, which is consistent with our assumptions and the naturalness of the breaking term.

Now that we know successful Leptogenesis is possible in this scenario, we must consider the various potential washout processes that must be prevented from destroying the generated asymmetry, deriving the associated parameter constraints to do so.

The chemical potential of the RH neutrino is solely sourced by the decay of the $\Phi_2$, whether the interaction $y \bar L  \Phi_2 \nu_R$ is in thermal equilibrium or not. The only requirement is that the lifetime of the $\Phi_2$ is short enough such that it decays while sphalerons are still in equilibrium. We consider the case of $ m_{\Phi_2} > T_{\textrm{EWPT}}\sim 100$ GeV, which approximately gives,
\begin{eqnarray}
\frac{y^2 m_{\Phi_2}}{8\pi} >  \sqrt{\frac{\pi^2 g_*}{90}} \frac{T_{\textrm{EWPT}}^2}{M_p}~,
\label{con1}
\end{eqnarray}
which requires $y > 2 \cdot 10^{-8}$ for $ m_{\Phi_2} \sim 1$ TeV scale.

We have necessitated that the $U(1)_X$ breaking term, $ \lambda_5 $, is out-of-equilibrium during the radiation epoch after reheating. This results in the corresponding condition,
\begin{eqnarray}
\frac{\lambda^2_5}{T^2} T^3 < H(T) \Rightarrow \lambda_5^2 \lesssim \frac{m_{\Phi_2}}{M_p}~.
\end{eqnarray}
which is easily satisfied as the correct baryon asymmetry requires $\lambda_5\sim 10^{-13}$.

If a Majorana mass is added for the RH neutrino, it must be out-of-equilibrium, with the rate for the Majorana mass mediation process being given by $y^4 m_{\nu_R}^2/T$ for $T > m_{\nu_R}$. Comparing this with the Hubble parameter $H$ at $T = T_{\textrm{EWPT}}$, we obtain the following approximate constraint, $y<10^{-3}\sqrt{\frac{1~\textrm{GeV}}{m_{\nu_R}}}~$. Thus, providing a large window of allowed Yukawa couplings for $m_{\Phi_2}> m_{\nu_R}$~.

Importantly, adding a Majorana mass for the RH neutrino will result in a tiny neutrino mass via radiative corrections \cite{Ma:2006km}. However, due to the smallness of the $\lambda_5$ required for successful Leptogenesis, such a mass term will be strongly suppressed.

The existence of the sequestered RH neutrinos will have observational implications for the number of radiation degrees of freedom in the universe, which will depend on the masses of $\Phi_2$ and $\nu_R$, as well as the size of its Yukawa interaction with the RH neutrino. This will be constrained by future probes of $\Delta  N_\textrm{eff}$ \cite{Li:2022yna}. Additionally, the asymmetric component in the RH sector may be a component of dark matter, depending on its mass, this will be a subject of future study.\\


\noindent \textbf{Discussion - }
We have presented a minimal renomalizable framework in which to achieve Dirac Leptogenesis via the Affleck-Dine mechanism, while also explaining the inflationary epoch. To do so, we have investigated the SM extended by a neutrinophillic Higgs doublet and a single RH neutrino. Here we have considered that the $ \Phi_2 $ does not have a finite vacuum expectation value, like the Inert Higgs Doublet scenario, to limit possible washout processes. While the RH neutrino has been assumed to be massless, enforced by a global $ U(1) $ lepton number symmetry, this setup has the potential to be embedded into various neutrino mass models, such as radiative neutrino mass model and a 2HDM, extending its potential phenomenological implications.

Beyond the successful explanation of the baryon asymmetry and inflation, this economical Leptogenesis mechanism exhibits a plethora of phenomenological implications enabling thorough testing. Depending on their mass, the remnant RH neutrinos may increase the number of relativistic degrees of freedom, and possibly act as a component of dark matter. The inflationary observables will be within reach of the upcoming LiteBIRD experiment \cite{Hazumi:2019lys} - alongside additional possible non-gausssianities and gravitational waves from violent preheating. Depending on the neutrino mass generating mechanism, there will also be unique lepton flavour violation signatures and possible non-standard neutrino interactions. At colliders, the new Higgs doublet leads to additional scalar signatures associated with the charged scalar $ H^+ $, alongside the neutral scalar $ H_0 $ and pseudoscalar $ A $ components, which have approximately degenerate masses ensured by the smallness of $ \lambda_5 $ term.\\


{\bf Acknowledgment.}---%
This work was supported by National Key R\&D Program of China under grant Nos. 2023YFA1606100. C. H. is supported by the Sun Yat-Sen University Science Foundation. NDB was supported by the Australian Research Council through the ARC Discovery Project DP210101636.

\bibliography{bibly2}

\end{document}